# Microsimulation of Space Time Trellis Code with Machine Learning

U. A. N. Ungku Chulan[1], M.T. Islam[2], N. Misran[3]
[1] Department of Electrical, Electronic and Systems Engineering, Universiti Kebangsaan Malaysia (UKM)

***Abstract*— Currently, the potential of microsimulation in reducing the temporal complexity of simulating space time trellis code has not been thoroughly ascertained. Therefore, this seminal work explores the possibility of using microsimulation in performing a pairwise comparison between competing generator matrices in code design. The validation of code construction is often done with simulation, which can be inherently time consuming. Microsimulation considerably cuts down the computational cost of simulation by employing smaller data and iteration. The effort is made feasible with the assistance of a machine learning model known as multilayer perceptron. When properly conducted, it can offer 93.86% accuracy and 98.25% reduction in temporal cost.**

## I. Introduction

The simulation [1] of Space Time Trellis Code (STTC) relies on randomization to model the behavior of a particular phenomenon. For instance, to simulate STTC in a Rayleigh fading channel, the random channel $h$ and noise $n$ are generated repeatedly to implement the simulation. This is vital in analyzing the average behavior of the BER vs SNR curve. Given this probabilistic paradigm, the accuracy of simulation is highly dependent on the number of samples employed. Larger samples often imply a better portrayal of the system in question. However, this comes at a price. More samples would traditionally instigate higher computational cost and time, which is far from desirable. Unlike simulation, the driving principle behind microsimulation is different altogether. Instead of utilizing a large sample size to achieve better accuracy, microsimulation seeks alternative ways of modeling the case of interest. For example, by reducing the number of samples or iterations needed [2]. This enables microsimulation to be significantly faster than simulation. Nevertheless, fewer samples normally increase the risk of error. In effect, an apropos recourse must be set in place. A promising strategy is to simplify the operations [2] involved such that less restrictive constraints are imposed. With regard to STTC, one of the most basic tasks in code design optimization is verifying the performance of the proposed code against those already established in the literature. This is often done by simulation, followed by a comparative analysis of their error curve performances. To illustrate this, consider two generator matrices $G_0$ and $G_1$ that are competing with one another (Figure 1). A series of iterations are performed for each of them and their results are calculated. By convention, the number of iteration is usually between 100 - 1000. Optimality is won by the generator matrix $G_\pi$ with the best result. In the case of $K$ optimal generator matrices $\{G_1 \dots G_K\}$ that are found in the literature and the best one is known ($G_\Omega$), it is hypothetically sufficient to do a comparison between $G_\alpha$ and $G_\Omega$ to know whether the newly proposed generator matrix $G_\alpha$ is more optimal than the rest. Simulating the performance of generator matrices can be a time consuming process. This is especially true when a vast quantity of generator matrix candidates is considered for optimization. Due to this, it is quite compelling to pursue the prospect of microsimulation instead.

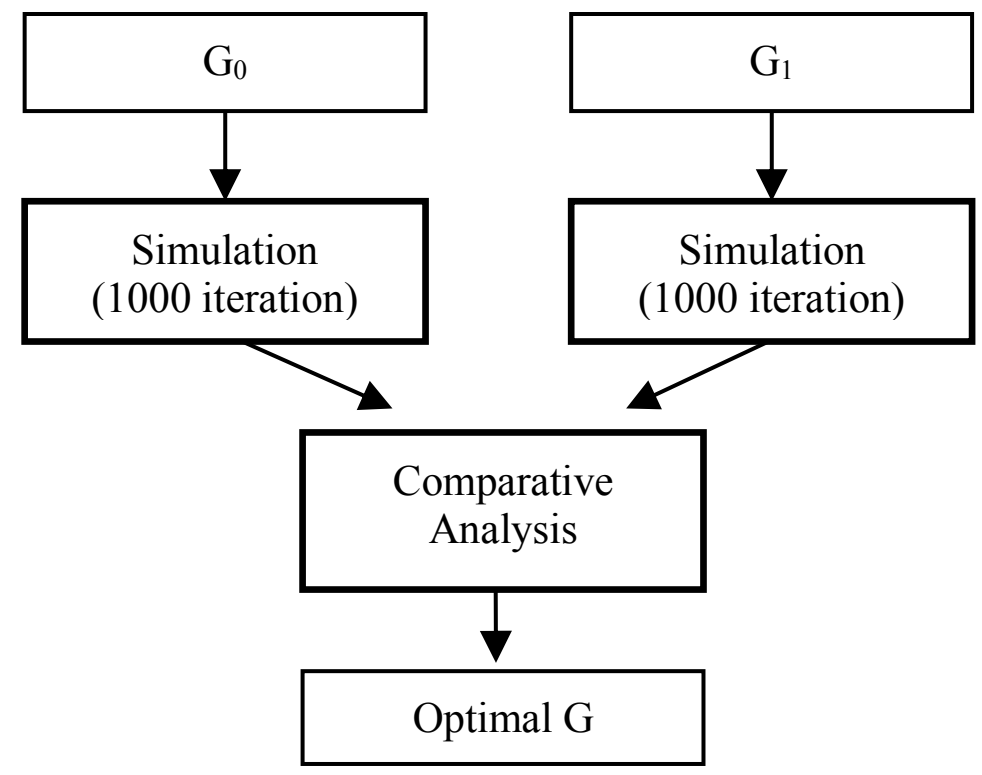

Fig. 1. Competition between two generator matrices via simulation

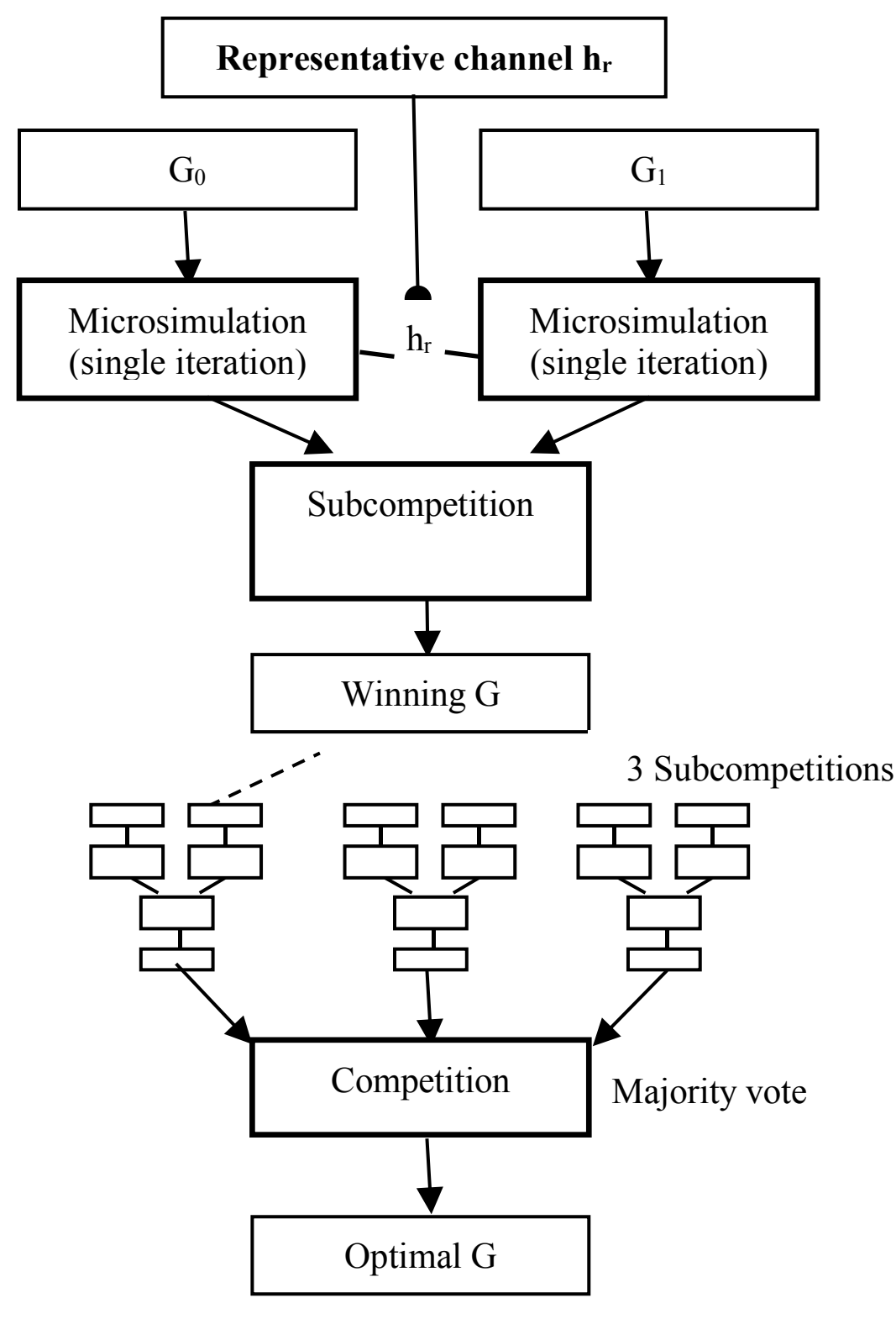

Fig. 2. Competition between two generator matrices via microsimulation

## II. System Model

The system model for this study is a Multi Input Multi Output (MIMO) system with two transmitters (*Tx=2)* and one receiver (*Rx=1)*. It is based on $y = hx + n$, which involve the transmitted signal $x$, channel matrix $h$, noise matrix $n$ and received signal $y$. Here, coding is achieved through STTC with QPSK modulation. The microsimulation in this study (Figure 2) proposes the use of representative channel matrix $h_r$ instead of a random one $h$. In Figure 3, a sample of the trellis diagram for the generator matrix $G^T$ = [0 0 2 1; 2 1 0 0] is given [3]. It is assumed that transmission is performed via the flat and slow Rayleigh fading channel. Apart from that, it must also be pointed out that the transmission data for the transmitted signal $x$ in microsimulation is not the same as simulation. Simulation normally requires 260 bits of random data as per the IS-136 standard [4]. Microsimulation on the other hand, uses the same data for transmission. This data is derived by concatenating all the possible variations of elementary input bits [5] for a particular modulation M. Thus, when M = 4, the transmission data or input bits $u$ = [00 00 01 10 11 00] including the preceding and succeeding zeros.

$$G^T = \begin{pmatrix} 0 & 0 & 2 & 1 \\ 2 & 1 & 0 & 0 \end{pmatrix}$$

| State | Output Symbols | | | |
|---|---|---|---|---|
| | **Input 00** | **Input 01** | **Input 10** | **Input 11** |
| **0** | 00 | 01 | 02 | 03 |
| **1** | 10 | 11 | 12 | 13 |
| **2** | 20 | 21 | 22 | 23 |
| **3** | 30 | 31 | 32 | 33 |

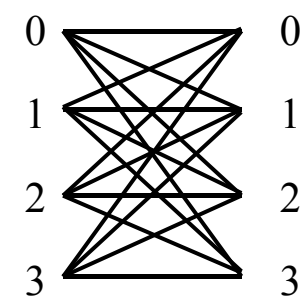

Fig. 3. Generator matrix and trellis diagram from Tarokh [8]

## III. MICROSIMULATION

As shown previously (Figure 1) the common practice of simulation [6] employs a profound number of iteration to work. This is the basis of monte carlo simulation where the randomization of a measure in large samples would enable the analysis for the average behavior of a system. Here, it is not difficult to hypothesize where the computationally expensive process actually lies. In reality, the number of iterations contributes significantly to the overall duration. In light of this, the microsimulation in this research addresses the problem from an entirely different approach (Figure 2). Instead of iterating the simulation for a substantial number of times (100 - 1000 iterations), only a single iteration is attempted. At first glance, this might appear rather extreme. However, the study borrows a concept known as representative sample from molecular biology [7,8]. It basically posits the existence of a few samples that can approximate the behavior of a large group of samples. In principle, instead of using a large size of random samples to capture the average behavior of the system under observation, a small size of representative samples is employed (Figure 4). The representative sample has the capacity to reflect the average behavior of the system, without the need for much greater sample size, as a random sample typically would. The conventional approach of finding a representative sample is often statistically inclined. However, this study introduces the use of machine learning to identify the best candidate as the representative sample.

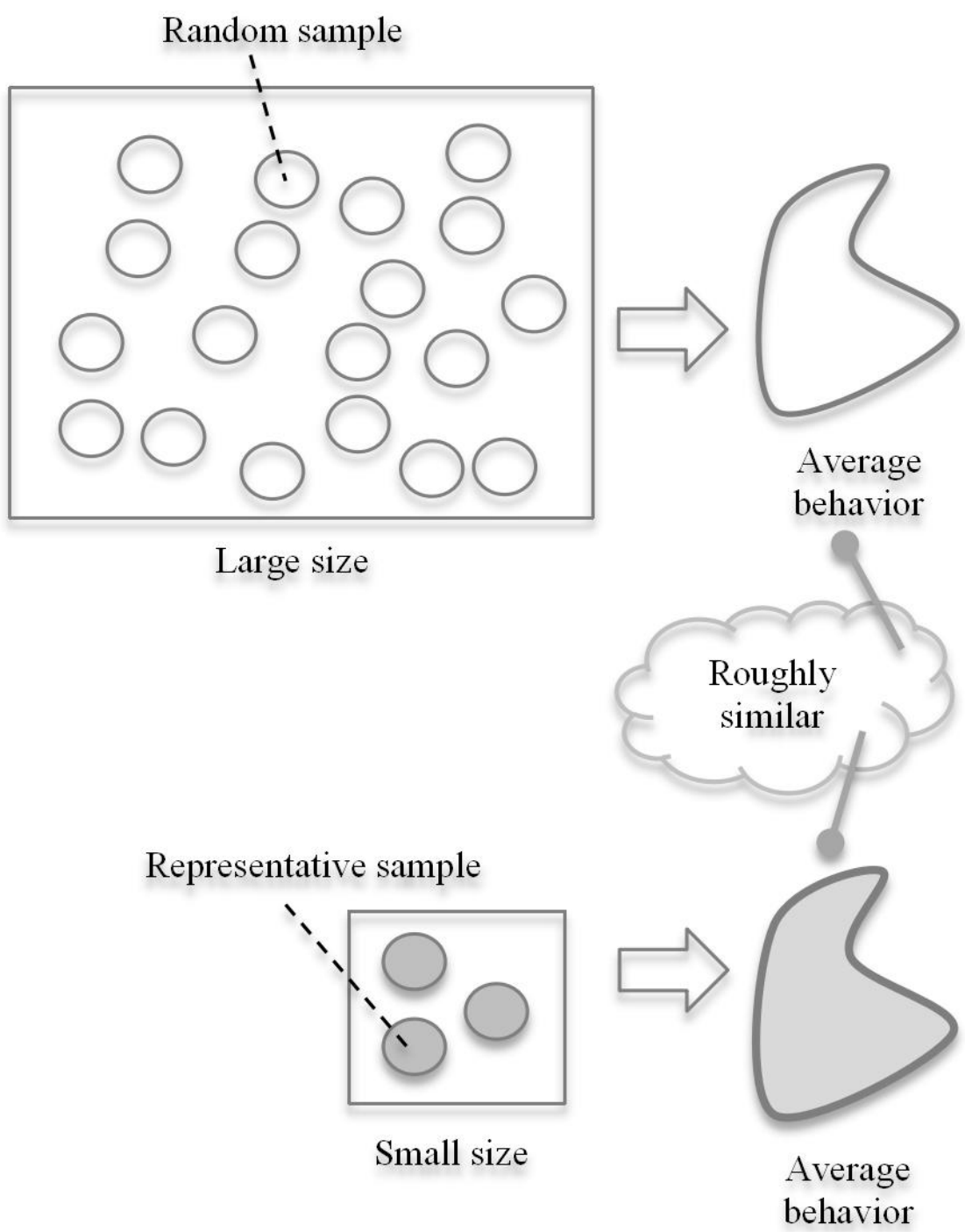

Fig. 4. Random sample vs representative sample

The idea of representative sample is not entirely a foreign idea in the simulation of STTC. To cite an example, for the prediction of irreducible error floor in STTC [5], elementary input bits are used as the representative sample instead of the common random input bits in simulation. To perform microsimulation (Algorithm I) on the competing generator matrix $G_0$ and $G_1$, a random channel matrix $h$ is generated. This is followed by a microsimulation competition between the two generator matrices with a representative channel matrix $h_r$. Here, the winner of the competition is deemed *as G_mic*. The features *D* from *G_mic* are then extracted and subsequently relayed to the machine learning model predict_with_MLP(*D*), which predicts whether the winner of the competition *G_mic* should be accepted or not. If it should (prediction == 1), then the random channel $h$ is elected as the representative channel $h_r$ and the winner is returned. Else, another random channel matrix $h$ is generated and the process repeats itself. Obviously, this is a form of lazy search as the effort is halted once the first solution is found. Additionally, the search is limited by the number of trials $T$ (total attempts made). Further elaboration on the machine learning technique is covered in the next section.

Algorithm I: Microsimulation

```
microsimulation(G0, G1)
   G = {}
   T = 100
   for k = 1 to T
      h = generate_random_channel_matrix( )
      G_mic = microsimulation_compete(G0, G1, h)
      D = feature_extraction(G_mic)
      prediction = predict_with_MLP(D)
      if prediction == 1
         G = G_mic
         break
      endif
   endfor
   return G
```

end

## IV. MACHINE LEARNING

The prediction of the representative channel matrix $h_r$ which occurs within *predict_with_MLP(D)* can be done with a machine learning model called multilayer perceptron (MLP) [9]. As for the architecture [10], the model contains one input layer, three hidden layers and one output layer. The number of nodes for the hidden layers are (10, 6, 5) respectively (Figure 5). Although numerous approaches are available in designing MLP, there is no definite way of deciding the total number of hidden layers as well as the number of nodes within them [11], as per the context demands. Context refers to the domain of which the model is applied. As such, the network is empirically constructed.

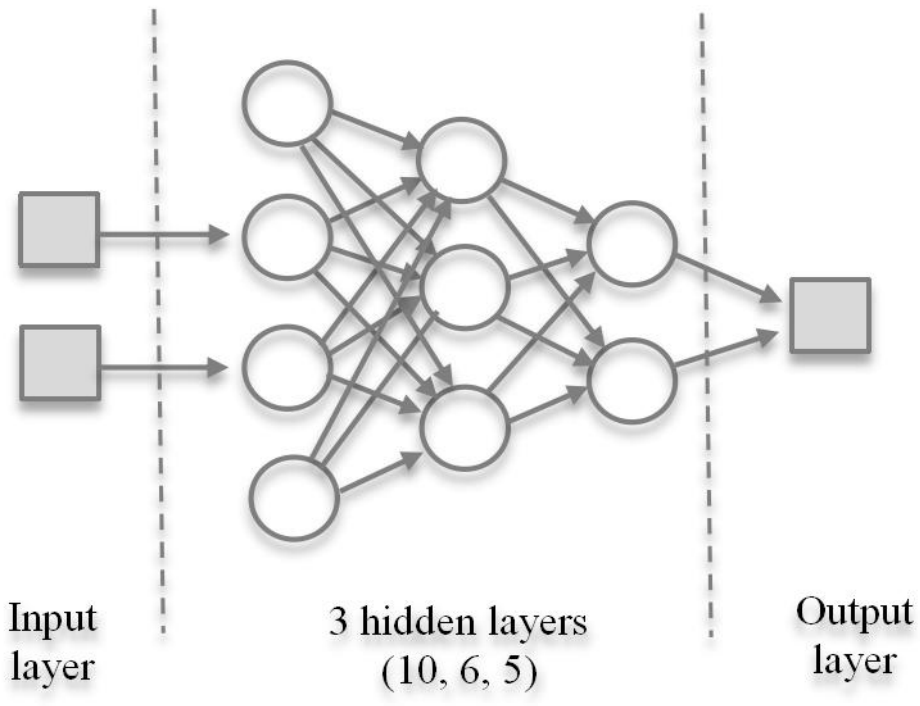

Fig. 5. Neural netwok architecture

For training, a 70 – 30 train-test split [12] is imposed on the data. Optimization is achieved with the LBFGS algorithm [13] and the maximum iteration is confined to 1000 to ensure proper convergence. In order to speed up the convergence process, the network relies on the ReLU activation function [14]. MLP is based on supervised learning. Thus, the right input-output pair must be provided. This is achieved by running a series of simulations to generate the required data for learning. Concerning feature extraction [15], the information gathered is enumerated in the table below (Table I). The result *R* of classification is determined by comparing the verdict of microsimulation against simulation. If they are agreeable, such that both microsimulation and simulation choose the same winner, then the classification result *R* is set to 1. However, if the microsimulation yields a different verdict from simulation, then the result *R* is set to 0.

TABLE I
FEATURE EXTRACTION

| FEATURE | CODE | DESCRIPTION |
|---|---|---|
| 1 | $G_0$ | GENERATOR MATRIX 0 |
| 2 | $G_1$ | GENERATOR MATRIX I |
| 3 | H | CHANNEL MATRIX |
| 4 | $N_0$ | AVERAGE NOISE ($G_0$) |
| 5 | $N_1$ | AVERAGE NOISE ($G_1$) |
| 6 | $B_0$ | AVERAGE BER ($G_0$) |
| 7 | $B_1$ | AVERAGE BER ($G_1$) |
| 8 | $Z_0$ | AVERAGE SNR WHERE BER REACHES ZERO ($G_0$) |
| 9 | $Z_1$ | AVERAGE SNR WHERE BER REACHES ZERO ($G_1$) |
| 10 | R | RESULT OF CLASSIFICATION |

**Data preparation** for the training phase entails a number of stages. To begin, datasets that correspond to two optimal generator matrices from the literature [3, 16], namely the ones from Tarokh [0 0 2 1; 2 1 0 0] and Baro [2 0 1 3 ; 2 2 0 1] are taken to generate the training data. Each optimal generator matrix is paired with 100 random generator matrices. Initially, a competition is performed via simulation to figure out the base of comparison where the actual winner is established. Afterwards, the competition is done via microsimulation. The process of preparing the microsimulation data is given in Algorithm II. $G_0$ signifies the optimal generator matrix that is chosen from the literature. $\{G_1, ..., G_N\}$ are the random and unique generator matrices. $G_0$ competes with $\{G_1 , ..., G_N\}$ where each competition is held 100 times with different random channel matrices [17]. The winner of microsimulation and simulation is *G_mic* and *G_sim* respectively. If the winners are congruent (*G_mic == G_sim*) then result $R = 1$. The extracted features *D* from a particular competition is updated with the result *R* and integrated with the overall data. Given that 100 competitions are held and each competition is repeated 100 times, a total of 10,000 data (100 x 100) is generated for each training dataset, giving a sum of 20,000 data for both training datasets.

Algorithm II : Data preparation

```
microsimulation_data(G0, {G1, .. , GN})
   data = {}
   for k=1 to N
     for i = 1 to 100
        hi = generate_channel_matrix()
        G_sim = simulation(G0, Gk)
        G_mic = microsimulation_compete(G0, Gk, hi)
         if G_mic == G_sim then
           result = 1
        else
           result = 0
        endif
        D = feature_extraction(G_mic)
        D(R) = result
        data = data ∪ D
     endfor
   endfor
   return data
end
```

In microsimulation, each competition (Algorithm III) between two generator matrices $G_0$ and $G_1$ would encompass three subcompetitions. Here, the generator matrix with most wins is selected as the winner $G_\pi$ via the majority vote algorithm [18]. The subcompetition (Algorithm IV) is based on three factors: SNR where BER reaches zero [19], average BER [20] and minimum BER [21]. A series of competition is facilitated by using one factor after another. In other words, if a factor fails, then the next one is enforced. First, a comparison is made in terms of the SNR of which the BER reaches zero, given by the ber_zero(G). The generator matrix that reaches zero the earliest is nominated as the winner. Here, select_G_with_min_value() denotes a function that returns the generator matrix with the minimal value. Supposed that the value for ber_zero($G_0$) is lower than ber_zero($G_1$), then select_G_with_min_value(ber_zero($G_0$), ber_zero($G_1$)) would return $G_0$. If the BER does not reach zero within the stipulated range of the SNR, then the value of ber_zero(G) = $\infty$. Now, if there is a tie, then a second comparison is executed in terms of the average BER. The generator matrix where the average BER is minimal would be the winner. Again, if there is tie, then a third comparison is done from the aspect of minimum BER. The generator matrix with the lowest minimum BER is the winner. Should there be a tie after the third comparison, then the winner is chosen randomly [22].

Algorithm III : Competition

```
microsimulation_compete(G0, G1, h)
   for k=1 to 3
      G(k) = microsimulation_subcompete(G0, G1, h)
   endfor
   G = majority_vote(G(1) .. G(3))
   return G
end
```

Algorithm IV : Subcompetition

```
microsimulation_subcompete(G0, G1,  h)
   winner = {}
   iteration       = 1
   ber(G0)         = simulation(G0, h, iteration)
   ber(G1)         = simulation(G1, h, iteration)
   ber_zero(G0) = get_snr_where_ber_reaches_zero(ber(G0))
   ber_zero(G1) = get_snr_where_ber_reaches_zero(ber(G1))
   ber_ave(G0)  = get_average(ber(G0))
   ber_ave(G1)  = get_average(ber(G1))
   ber_min(G0)  = get_mininum(ber(G0))
   ber_min(G1)  = get_minimum(ber(G1))
   if ber_zero (G0) != ber_zero(G1) then
      winner = select_G_of_min_value(ber_zero(G1), ber_zero(G1))
   else
      if ber_ave(G0) != ber_ave(G1)
         winner = select_G_of_min_value(ber_ave(G0), ber_ave(G1))
      else
         if ber_min(G0) != ber_min(G1)
            winner = select_G_of_min_value(ber_min(G0), ber_min(G1))
         else
            winner = random(G0, G1)
         endif
      endif
   endif
   return winner
end
```

## V. EXPERIMENTATION

The experimentation parameters are given in Table II. MATLAB is used to develop the STTC portion of the code while Python and scikit-learn [23] are leveraged to implement the machine learning part. To observe the extent of generalizability, the research employs completely different dataset of generator matrices for training and testing. This is purposely done to investigate whether the approach is afflicted by overfitting. If the method can maintain reasonably high accuracy despite the different datasets, then it does not succumb to overfitting. Two optimal generator matrices are taken from the literature for training while seven more for testing (Table III). Each optimal generator matrix is competed against 100 random and unique generator matrices. Competition is first performed with simulation and then with microsimulation. The former provides a benchmark for the latter. Furthermore, two forms of microsimulation are carried out: microsimulation and microsimulation+MLP to ascertain the impact of integrating MLP into the approach. The mean accuracy of microsimulation and microsimulation+MLP is 0.5914 and 0.9386 respectively (Table IV). This implies that MLP affords an improvement of 62.11% for accuracy. Upon closer examination, it is also discovered that the accuracy of microsimulation+MLP is more stable than its counterpart. This can be attributed to the use of the representative channel instead of a random one where the standard deviation (s) and variance ($s^2$) of microsimulation+MLP are both significantly lower than microsimulation. From the aspect of performance (Table V), microsimulation+MLP allows a reduction of temporal cost from 82.0672s to 1.4382s when compared with simulation. This denotes a staggering improvement of 98.25%. In retrospect, this is to be expected. Microsimulation employs a much smaller input bits and number of iterations when compared to simulation. While simulation relies on 260 bits of random input data and 100 - 1000 iterations, microsimulation requires only 12 bits of identical input data and 1 iteration per subcompetition. Nevertheless, the performance of microsimulation+MLP is not as stable simulation. Exhibiting a comparatively higher standard deviation (s) at 0.9137, it requires at least 0.3259 seconds and at most 2.5734 seconds for processing. This includes the time of searching for the representative channel matrix $h_r$. It is reasoned that the duration instability is caused by the random traversal of the search.

TABLE II
EXPERIMENTATION PARAMETERS

| Aspect | Parameter | Value |
|---|---|---|
| Machine | Processor | Core2Duo E8400 @ 3.00 GHz |
| | RAM | 8Gb |
| Software | MATLAB | version R2015a (8.5.0) |
| | Python | version 2.7.15 |
| | Scikit-learn | version 0.20.4 |
| System Model | Transmitter | 2 |
| | Receiver | 1 |
| | SNR Range | [0, 24] |
| | Channel | Flat & Slow Rayleigh fading |
| | Modulation | 4 (QPSK) |
| STTC | Total State | 4 |
| | Generator Matrix | 4x2 |
| Data | Simulation | 260 bits (random) |
| | Microsimulation | 12 bits (constant) |
| | Total Sample | 707 generator matrix G |
| | Total case | 7 cases (7 x 101 G) |
| | Each case | (1 optimal G ) vs (100 random G) |

TABLE III
DATASET FOR EXPERIMENTATION

| CASE | NAME | GENERATOR MATRIX $^T$ |
|---|---|---|
| 1 | BANARJEE [1] | [ 0 0 1 2 ; 1 2 0 0 ] |
| 2 | ILHAN [24] | [ 2 3 2 1 ; 2 3 0 2 ] |
| 3 | CHEN [25] | [ 0 2 1 0 ; 2 2 0 1 ] |
| 4 | INOUE [26] | [ 2 3 2 0 ; 0 2 1 2 ] |
| 5 | YAN [27] | [ 2 0 1 2 ; 2 2 2 1 ] |
| 6 | LIAO [28] | [ 0 0 1 2 ; 2 1 0 0 ] |
| 7 | HONG [29] | [ 0 2 2 3 ; 2 2 1 2 ] |

TABLE IV
ACCURACY OF MICROSIMULATION

| CASE | MICRO SIMULATION | MICRO SIMULATION +MLP | IMPROVEMENT (%) |
|---|---|---|---|
| 1 | 0.5800 | 0.9500 | 63.79 |
| 2 | 0.6500 | 0.9900 | 52.30 |
| 3 | 0.4900 | 0.9600 | 95.91 |
| 4 | 0.6600 | 0.9500 | 43.94 |

| 5 | 0.5200 | 0.8800 | 69.23 |
|---|---|---|---|
| 6 | 0.7400 | 0.9000 | 21.62 |
| 7 | 0.5000 | 0.9400 | 88.00 |
| **MEAN** | **0.5914** | **0.9386** | **62.11** |
| S | 0.0949 | 0.0372 | - |
| $S^2$ | 0.0090 | 0.0014 | - |

TABLE V
PERFORMANCE OF MICROSIMULATION

| CASE | SIMULATION (SECOND) | MICRO SIMULATION +MLP (SECOND) | IMPROVEMENT (%) |
|---|---|---|---|
| 1 | 81.3207 | 0.4845 | 99.40 |
| 2 | 83.0385 | 1.8879 | 97.73 |
| 3 | 81.7811 | 2.5734 | 96.85 |
| 4 | 81.9344 | 1.8822 | 97.70 |
| 5 | 82.1167 | 0.6907 | 99.16 |
| 6 | 82.2535 | 0.3259 | 99.60 |
| 7 | 82.0258 | 2.2226 | 97.29 |
| **MEAN** | **82.0672** | **1.4382** | **98.25** |
| S | 0.5227 | 0.9137 | - |
| $S^2$ | 0.2732 | 0.8348 | - |

Regarding the visible impact of the random channel matrix $h$ vs representative channel matrix $h_r$, illustrations of the results can be seen in Figure 6 and Figure 7 respectively. To demonstrate the impact of representative channel, the optimal generator matrix $G_0$ = [0 2 2 3; 2 2 1 2] from Hong [29] competes with a random generator matrix $G_1$ = [2 2 0 1; 0 0 2 2]. Now, the focus here is the extent of discernibility between the error curves of the generator matrices when deciding the winner. In other words, on how distinct the error curves are between the competing generator matrices such that a verdict can be given without uncertain deliberation. A channel matrix is deemed favorable when it is capable of making a clear distinction between the generator matrices in question. As shown, when a random channel matrix $h$ is used (Figure 6), it is difficult to discern the winner as the error curves overlap. Contrast this with the second case (Figure 7). Only a single iteration is carried out and yet, the difference between the error curves of $G_0$ and $G_1$ is quite distinct. This distinction allows microsimulation to decide the winner of the competition more effectively. There is a limitation to the approach. Microsimulation lacks the granularity of simulation when it comes to detailing the error curve. For instance, it does not convey the actual BER of a particular SNR for the error curve. However, it manages to distinguish the overall BER-vs-SNR performances between the generator matrices at hand.

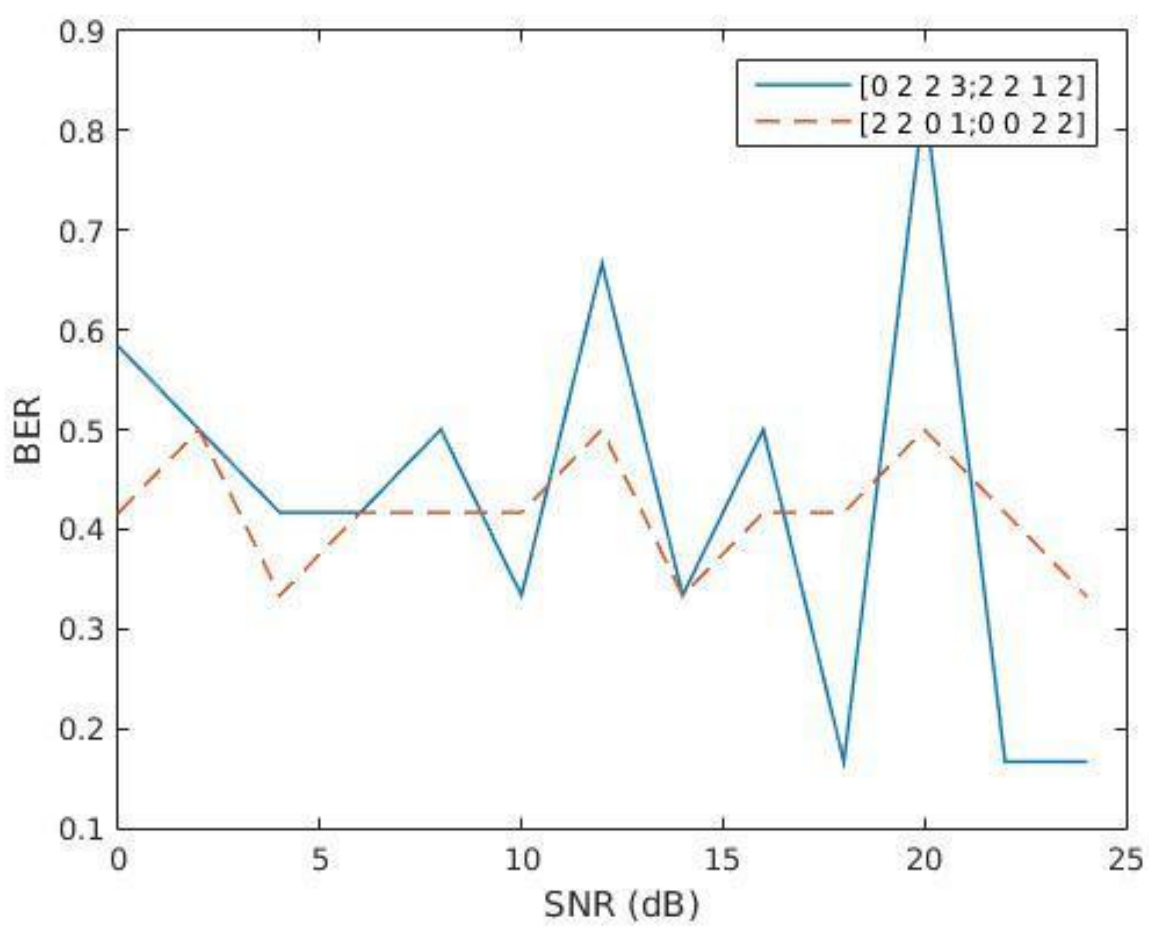

Fig. 6. Overlapping error curves in microsimulation (single iteration) with random channel matrix $h$

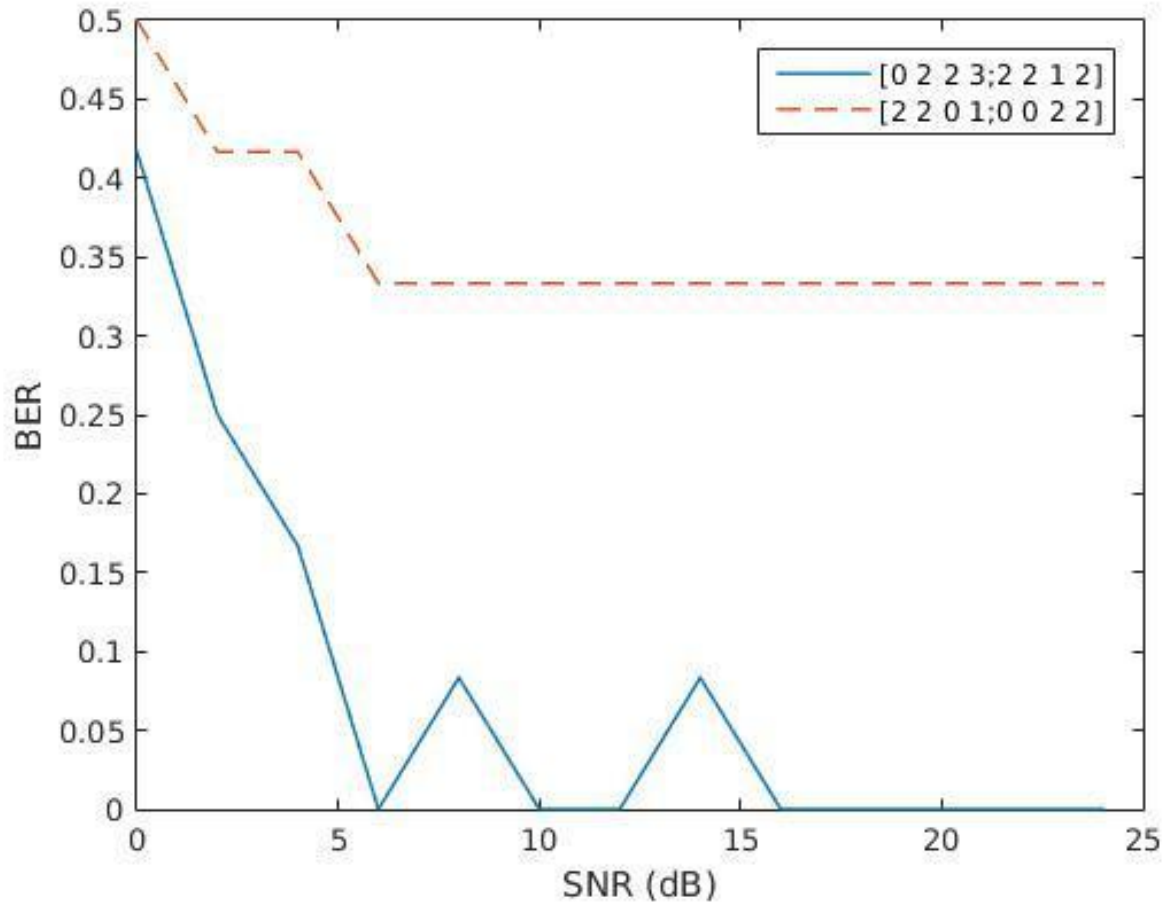

Fig. 7. Distinct error curves in microsimulation+MLP (single iteration) with representative channel matrix $h_r$

## VI. CONCLUSION

Microsimulation is a viable alternative in conducting a pairwise comparison between competing generator matrices in the effort of optimizing the code design of STTC. It can reduce the cost of simulation by approximately 98% and still achieve roughly 94% accuracy. This is feasible with the usage of a representative channel $h_r$ that is found with multilayer perceptron (MLP). The study however, is conducted for the core system model of STTC. Thus, further effort must be invested on more complex ones in the future.